\documentclass[onecolumn,pre,preprint]{revtex4}
\usepackage{graphicx}
\usepackage{dcolumn}
\usepackage{amsmath}
\usepackage{times}
\usepackage{bm}

\begin{document}
\title{Onset of slow dynamics in difluorotetrachloroethane
glassy crystal}
\author{F.~Affouard, E.~Cochin, F.~Dan\`ede, R.~Decressain, M.~Descamps}
\affiliation{
Laboratoire de Dynamique et Structure des Mat\'eriaux Mol\'eculaires,\\
CNRS UMR 8024,
Universit\'e Lille I,\\
59655 Villeneuve d'Ascq Cedex France}
\author{W.~Haeussler}
\affiliation{
Institut Laue-Langevin, 38042 Grenoble, France
}
\begin{abstract}
Complementary Neutron Spin Echo and X-ray experiments and Molecular Dynamics simulations have
been performed on difluorotetrachloroethane (CFCl2-CFCl2) glassy crystal.
Static, single-molecule reorientational dynamics and collective dynamics properties are investigated.
The orientational disorder is characterized at different temperatures and a change in nature 
of rotational dynamics is observed.
We show that 
dynamics can be described by some scaling predictions of the Mode Coupling Theory (MCT)
and a critical temperature $T_{c}$ is determined.
Our results also confirm the strong analogy between molecular liquids and plastic crystals
for which $\alpha$-relaxation times and
non-ergodicity parameters are controlled by the non trivial static correlations 
as predicted by MCT.
\end{abstract}

\maketitle

\section{Introduction}
Understanding of the mechanisms leading to the glass transition 
has been a subject of intense research 
in recent years and it is 
still a matter of numerous theoretical and experimental studies~\cite{Debenedetti_nature01}.
A fundamental question particularly concerns the microscopic description of the precusor cooperative mechanisms
which develop in the high temperatures pico-nanosecond regime (ps-ns) far above the glass transition temperature $T_{g}$.
It is now commonly suspected, mainly from theoretical and numerical results, that dynamical behaviours of glass-forming systems 
could be described by involving
two characteristic temperatures~\cite{Sastry_nature01}:
i) $T_{A}$
($T_{S}$ or $T_{x}$ are also seen) 
corresponding 
to the onset of slow dynamics
where
non-exponential relaxation and non-Arrhenius behavior emerge~\cite{Sastry_nature01}. First appearance of
rotational heterogeneity are also reported~\cite{Kim_pre03}.
The role of $T_{A}$ in the vitrication process is still not completely understood but 
it is often considered as the temperature below which dynamics are influenced by the
potential energy landscape~\cite{Sastry_nature01}.
ii) $T_{c}$, the critical temperature predicted by the Mode Coupling Theory 
MCT~\cite{Goetze}.
In the last two decades, enormous work has been devoted to this theory and it is so far
the only one which provides a
microscopic description of simple liquids. 
The intrinsic basis of MCT states that
the behaviour of any time-dependent correlators describing the dynamics of the system is only controlled
by its static density correlator \emph{i.e} $S(Q)$ and  predicts the existence of a
critical temperature $T_{c}$
indicating a bifurcation from ergodic to non-ergodic dynamics.
The role of $S(Q)$ and 
the scaling properties predicted by MCT approaching
$T_{c}$ have been successfully validated in numerous experiments and molecular dynamics (MD) simulations~\cite{Schilling_03}.
However, 
no sharp transition is observed at $T_{c}$ itself owing some activated processes not included in the theory
which restores ergodicity. Thus,  $T_{c}$ 
is also associated to a dynamical crossover
to a 'landscape-dominated'
regime in which diffusion processes can be described in terms of
thermally activated jumps.

Most of the glassformers are molecular liquids and accordingly possess
both translational and rotational degrees of freedom (TDOF and ODOF).
Approaching the glass transtion, the coupling between TDOF and ODOF is far from being understood
and the precise role of the ODOF remains unclear as demonstrated by the so-called low-temperature
translation-rotation paradox~\cite{Heuberger_jpc96,Blackburn_jpc96}.
MCT has been initially developped to provide
a microscopic understanding of simple atomic liquids. However, general predictions
have also been successfully validated for molecular compounds.
Recently, some extensions of MCT called molecular mode coupling theory (MMCT)~\cite{Schilling_03}
have been proposed to take ODOF into account:
one diatomic probe molecule in an atomic liquid~\cite{Chong_pre01}, liquids made of linear molecules~\cite{Schilling_pre97} or
water~\cite{Fabbian_pre99} to cite only a few. The authors have particularly shown that some of the basic
predictions of MCT still hold owing TDOF/ODOF coupling. As revealed by
recent MD investigations~\cite{Chong_epl03,Chong_pre04} performed on orthoterphenyl (OTP),
coupling of the rotational dynamics to the center-of-mass motion can be complex.

An alternative to
the glassforming molecular liquids is offered by
at least two partially disordered systems which provide
an interesting solution in order to investigate only ODOF during the freezing process :
                                                                                                                                               
i) The so-called \emph{orientational glasses}
such as $\mathrm{Ar_{1-x}(N_{2})_{x}}$  or
$\mathrm{(KCN)_{x}(KBr)_{1-x}}$ mixed crystal where rotational motions are frustated by a quenched disorder
externally imposed
by a dilution~\cite{Hochli}
K.~H.~Michel and coworkers~\cite{Michel2}
developped a MCT for these systems
which involved the interplay or
rotational-translational and rotational-strains
coupling where strains result from the substitutional
disorder.
                                                                                                                                               
ii) \emph{Glassy crystals}~\cite{Suga_jncs74}
such as cyanoadamantane~\cite{Affouard_jncs02}, ethanol~\cite{Criado,Benkhof_jpcm98},
orthocarborane~\cite{Winterlich_prl03} or
cyclooctanol~\cite{Brand_prb97} which are true rotational analogs of canonical liquid glassformers.
These systems are
generally composed of globular molecules and exhibit a
partially disordered phase, called plastic,
which possesses a crystalline translational order
and a dynamical orientational disorder. They
can be rapidly cooled from the plastic phase and
exhibit
a step of the specific heat at the calorimetric temperature $T_{g}$
or a non-Arrhenius behavior
of the rotational relaxation times.

No microscopic theory has been developped for plastic crystals so far. However,
in~\cite{Affouard_epl01,Affouard_jpcm01,Affouard_prl01},
we have particularly shown from NMR and Raman experiments, and
MD computer simulations that
some predictions of
the idealized version of the MCT (critical temperature $T_{c}$ and time scaling laws)
were
able to describe
relatively well rotational dynamics of different plastic crystals.
It should be noted that the other remarkable temperature $T_{A}$ was also identified. 
These intriguing previous results call for new investigations
to clarify the similarity between slow dynamics behaviour of plastic crystals and  molecular liquids
and in particular to discuss the role of the static correlations $S(Q)$.
In this paper, we present for the first time results of
Neutron spin-echo (NSE) and X-ray experiments
and MD simulations of the
glassy crystal difluorotetrachloroethane (DFTCE).
This compound is composed of simple molecules $\mathrm{CFCl_{2}-CFCl_{2}}$ close to 
dumbbells extensively used in MD calculations as prototype of molecular glass-formers liquids~\cite{Kaemmerer_pre97,Michele_pre01,Chong_pre02}. 
DFTCE has been experimentally widely studied
and presents a rich variety of unusual properties. DFTCE exhibits a glass transition of the overall rotation of the 
molecules at $T_{g} = 86 $ K~\cite{Kishimoto_bcJapan78}.
Two 
additional heat anomalies are found at 60 and 130 K and associated respectively to 
a sub-$T_{g}$ $\beta$ process and the freezing of the transformation between \emph{trans} and \emph{gauche} 
conformation of the molecule. 
Changes concerning the nature of dynamics in this system
has been reported from NMR experiments~\cite{Stokes_jcp79}, Brillouin and 
dielectric spectroscopy~\cite{Kruger_jpcm94}. In this latter investigation, 
a crossover from individual to correlated rotational motions was particularly suggested.

\section{Experiments and details of the simulation}

The samples used in both X-ray and NSE experiments consist 
of commercial 1,2-difluorotetrachloroethane purchased from Sigma-Aldrich
and purified by melting zone.

The X-ray diffraction experiment was performed on a laboratory
diffractometer equipped with an INEL curved sensitive detector CPS 120 made up of 
4096 channels and allowing us to record simultaneously X-ray diffraction on a $2\theta$ range
of about 120 deg. The incident beam was monochromatized with a bent of quartz crystal which selects
the $\mathrm{K_{\alpha_{1}}}$ wavelength of a Cu X-ray tube ($\lambda = 1.54056 $ \AA). Powder DFTCE was 
introduced in Lindemann glass capillary of 0.5 mm in diameter. It was mounted on the axis of the 
diffractometer and rotated during the experiment
in order to reduce the effect of preferential orientation. The correspondence between the channel number 
and the $2\theta $ angle was determined by a preliminary X-ray experiment of cubic 
$\mathrm{Na_{2}Ca_{2}Al_{2}F_{14}}$ (NAC)~\cite{Evain93} 
with a cubic spline interpolation between the Bragg peaks of NAC. 
Low temperatures were achieved by placing
the sample into a gaseous
nitrogen flow at the required temperature. Using this procedure,
the sample is quenched within a
few seconds, and its temperature can be controlled within 0.5 K between 110 K and 300 K.
In the present experiment, the X-ray pattern was recorded at 140 K during 3600 s for a sample quenched at 2K/min.

The NSE experiments were performed on the IN11 spectrometer
at the Institut Laue-Langevin (ILL), Grenoble, France.
We used the multidetector version of IN11
which allows simultaneous measurements at
different scattering angles. 
A setup with an incident wavelength of $\lambda = 5.5 \ \mathrm{\AA}$ 
($\Delta \lambda/\lambda \simeq 16 $ \%)
gave access to
wave numbers approximately between 1.28 and $1.72 \ \mathrm{\AA^{-1}}$
and we covered the time range 8 ps - 1 ns.
We made measurements of the coherent signal of 
poly-crystalline systems filled into a flat Al container
from $T = $ 130 to 220 K in steps of 10 K.
For normalization, a resolution scan of the elastic scattering was realized at $T = 5 $ K
in the glassy phase
and all data were divided by the polarization at this temperature.
Two samples of different mass identified as A ($m \simeq 0.33 $ g) and B ($m \simeq 2 $ g)  were investigated.
Although preliminary measurements showed that transmission of sample A was better,  93\% to be compared to 77\% obtained
for sample B,
this latter was prefered for most of the measurements since 
its scattering signal was more intense. Unfortunately, only data obtained from sample A are available
at $T = $ 130 and 140 K.   

MD calculations were performed on a
system of $N = 686$ ($7 \times 7 \times 7$ bcc crystalline cells)
molecules. Each DFTCE molecule  $\mathrm{CFCl_{2}-CFCl_{2}}$
is described by its 8 atoms and considered as
a rigid unit.
They interact through a Buckingham short range atom-atom potential
(see parameters in table~\ref{table1})
and the electrostatic contributions have been neglected
since the DFTCE molecule possesses a weak dipolar moment.
No structural or dynamical change has been found for a system
where electrostatic interactions
are taken into account.
It should be noted that
both structure and dynamics were found in good agreement
with experimental results (see figures~\ref{figure1} and~\ref{figure2}).
Newton's equations of motion were solved
with a time step of $\Delta t = 5$ fs. We worked
in the NPT statistical ensemble
with periodic boundaries conditions where the simulation
box is allowed to change in size and shape.
MD simulations were done
for
a sample corresponding to the DFTCE orientationally disordered phase
at 14 different temperatures from
$T =$ 130 to 260 K in steps of 10 K.  It should be mentioned that
our very simple DFTCE model allows us to perform
very long MD runs of about 50 ns.

In this paper,
single-molecule reorientational dynamics of DFTCE
will be obtained
from
the
first two self angular correlation functions $C_{l=1,2}$~\cite{Kaemmerer_pre97} defined as:
\begin{equation}
C_{l}(t) =
\frac{1}{N}\sum_{i=1,N}
\left \langle
P_{l} \left(
\vec{u}_{i}(t).\vec{u}_{i}(0)
\right)
\right \rangle
\label{PLegendre}
\end{equation}
where $P_{l}$ is the $l$-order Legendre polynomial and $\vec{u}_{i}$ is a 
unit vector directed along the $\mathrm{C_{3}}$-symmetry carbon-carbon molecular axis of the 
DFTCE molecule $i$ (see Fig.~\ref{figure1}).
From $C_{l=1,2}$, we can obtain clarification on
the nature of dynamical changes and anisotropy of the molecular
motions.
Experimentally,
$C_{1}(t)$ can be directly measured in dielectric relaxation and
$C_{2}(t)$ in Raman scattering~\cite{Sherwood}.
$C_{l=1,2}$ can also be related to
the informations obtained from NMR relaxation measurements~\cite{Affouard_epl01}.

In order to investigate collective dynamics, the intermediate scattering function as it can be classically obtained from coherent neutron
scattering experiments will be mainly considered:
\begin{equation}
S(\vec{Q},t)= \langle
 \rho_{\vec{Q}}(t)\rho_{\vec{Q}}(0) \rangle \label{Scoh}
\end{equation}
where $\rho_{\vec{Q}}(t)$ is the time-dependent density correlator:
\begin{equation}
\rho_{\vec{Q}}(t) = \sum_{\alpha}b_{\alpha}exp[i\vec{Q}.\vec{r}_{\alpha}(t)]
\label{rho}
\end{equation}
where the sum is over all the atoms $\alpha$ of the system.
$b_{\alpha}$ and $\vec{r}_{\alpha}$ are
the coherent scattering length
and the position of the $\alpha$ atom respectively.
An average over isotropically distributed
$Q$-vectors having the same modulus $Q$
is performed in order to obtain $S(Q,t)$ for a polycrystalline sample. In general,
$S(Q,t)$ can be expressed as~\cite{Dolling_mp79}:
\begin{equation}
S(Q,t)= S_{c}(Q) + S_{d}(Q,t)  \label{ScohDecomp}
\end{equation}
where $S_{c}(Q) = |\langle \rho_{\vec{Q}} \rangle |^{2}$ is the coherent elastic scattering
and
\begin{equation}
S_{d}(Q,t)= \langle \delta \rho_{\vec{Q}}(t) \delta \rho_{\vec{Q}}(0)  \rangle  \label{Sd}
\end{equation}
where $\delta \rho_{\vec{Q}}(t) =  \rho_{\vec{Q}}(t) - \langle \rho_{\vec{Q}} \rangle$
is the fluctuation of the time-dependent density operator.
For plastic crystals,
$S(Q,t)$ is identical to $S_{d}(Q,t)$ except for $Q$-vectors
corresponding to Bragg peaks
for which the long time limit of $S(Q,t)$ reaches
the non-zero value $S_{c}(Q)$.
This behaviour
is found at all temperatures and it is associated to the crystalline order of the TDOF and not the
freezing process of the ODOF.
Similarly to the intermediate scattering function,
the static correlator can be decomposed
into two components $S(Q)= S_{c}(Q) + S_{d}(Q)$
where $S_{d}(Q)$ is the diffuse scattering:
\begin{equation}
S_{d}(Q)=  \langle |\delta\rho_{\vec{Q}}|^{2} \rangle
= \langle |\rho_{\vec{Q}}| ^{2}  \rangle - |\langle \rho_{\vec{Q}} \rangle |^{2}
\end{equation}

\section{Results and discussion}
\subsection{Static}
\label{static}
Owing the high rotational disorder, plastics crystals yield only a few diffraction peaks
but exhibit intense and highly structured diffusive scattering which provides useful information to
characterize rotational motions and most probable orientations~\cite{Dolling_mp79}.
Fig.~\ref{figure1} shows both numerical and experimental static structure factors obtained 
in the $Q$-range $[0.3 - 6.5]$ \AA$^{-1}$.
The diffraction pattern of the plastic phase presents eight Bragg peaks localized at 
1.272, 1.799, 2.203, 2.543, 2.844, 3.114, 3.596 and 3.816 \AA$^{-1}$
which are well indexed  as a bcc structure
having a cell parameter of 6.985 \AA$^{-1}$. The cell parameter is found 2\% smaller from MD simulation.
This result confirms the preliminary X-ray diffraction investigation of this 
crystal performed by Kishimoto~\emph{et Al}~\cite{Kishimoto_bcJapan78}.
The elastic and the diffuse scattering have been extracted from the MD study and are found in good agreement with the 
X-ray experiment: maxima of $S_{c}(Q)$ coincide with the position of the experimental Bragg peaks and  
maxima of $S_{d}(Q)$ with the broad bumps of the diffuse scattering at $Q \simeq $ 1.3, 2.34, 3.36 and 4.8 \AA$^{-1}$. 
It should be mentionned that an additional diffuse scattering bump is observed experimentally close to $Q \simeq 1.6 $ \AA$^{-1}$. This feature
which is also present in the pattern of the liquid phase (see Fig.~\ref{figure1}) mainly derives from the scattering contribtuon of the glass capillary sample holder.
Moreover,
a strong similartiy of the X-ray pattern is observed between the liquid and plastic phase.
Such remarkable feature has been already reported for
ethanol~\cite{Fayos_prl96} or
$\mathrm{CBr_{4}}$~\cite{Dolling_mp79}.
The first sharp diffraction peak of the liquid particularly corresponds to the first and most intense Bragg peak of the plastic phase.
Broad bumps of the diffuse scattering are found almost identical in both phases and
merge at the largest $Q$-vectors
since the same intramolecular structure is probed.
This result could suggest a strong analogy between
the supercooled liquid and the supercooled plastic crystal of the same compound
in the framework of the MCT which predict that dynamics are controlled by non trivial static correlations. This question will be
 adressed in an other
paper~\cite{Affouard_prl04}.

The orientational distribution of the DFTCE molecules can be described with the probability distribution function $f(\vec{u})$ 
where $\vec{u} = (x,y,z)$
is a unit vector along the carbon-carbon molecular axis of DFTCE (see Fig.~\ref{figure1}). 
Since the site symmetry of the crytal is $O_{h}$, 
$f(\vec{u})$ may be expanded as a series of cubic harmonics $K_{i}$~\cite{Chelli_jcp97}: 
\begin{equation}
f(\vec{u}) = \frac{1}{4\pi} \left ( 1 + \sum_{i=4,6,8,...}\langle K_{i} \rangle K_{i}  \right )
\end{equation}
where $K_{4} = (21/16)^{1/2}(5Q-3)$, $K_{6} = (13/128)^{1/2}(462R + 21Q -17)$ 
and $K_{8} = (561/1024)^{1/2}(65Q^{2} -208R -94Q +33)$
with $Q=x^{4}+y^{4}+z^{4}$
and $R=x^{2}y^{2}z^{2}$. The $\langle K_{i} \rangle$ coefficients are the averages of the corresponding $K_{i}$.  $K_{4}$, $K_{6}$ and $K_{8}$ have been 
calculated from MD simulations at different temperatures and are reported in table~\ref{table2}. This table clearly shows no strong ordering of the DFTCE molecules along 
the prefered [111], [110] or [100] directions of the cubic lattice consistent with the high rotational disorder of 
plastic phases. However, a slight preference for [111] directions is observed. Upon lowering the temperature,
we also found an increase of the orientational probability along [111] unlikely to [110] or [100].
These results have been also confirmed from different snapshots of individual molecule trajectories as displayed in Fig~\ref{figure_snap}.

\subsection{Single-molecule reorientational dynamics}
Single molecule reorientational dynamics have been investigated from the self angular correlation 
functions $C_{l=1,2}$ defined in Eq.~\ref{PLegendre}. 
Figure~\ref{figure_snap} shows $C_{2}(t)$ time correlation
functions for all investigated temperatures from 130 K to 260 K.
$C_{1}(t)$ time correlation
functions, not displayed in the present paper, shows a very similar
behaviour to the $C_{2}(t)$ functions.
At high temperatures, $C_{2}(t)$ rapidly decays zero and can be described
by a simple exponential shape.
When lowering
the temperature, dynamics slow down and
a two-step process is seen as already observed in chloroadamantane
plastic crystals~\cite{Affouard_epl01,Affouard_prl01} and
in supercooled liquids~\cite{Kaemmerer_pre97}.
This process is often interpreted 
in term of $\alpha - \beta$ 
relaxations which will be carefully analyzed in the next section for collective dynamics
 in the framework of MCT.
Reorientation times $\tau_{1}$ and $\tau_{2}$ have been extracted 
from the time it takes for their respective self angular correlation functions $C_{l=1,2}$
to decay $e^{-1}$ of their initial values.
The ratio $\tau_{1}/\tau_{2}$ 
is displayed in Fig.~\ref{figure_snap}. 
For uncorrelated small angular steps motion, it can be
shown that $C_{l}(t)$ functions follow
an exponential decay
$exp[-t/\tau_{l}]$
where $1/\tau_{l} = l(l+1)D_{r}$ and $D_{r}$ is the
rotational diffusion coefficient. 
At high temperatures, $\tau_{1}/\tau_{2}$ is close to the value of 3 in good agreement 
with this simple diffusion model. 
Upon lowering the temperature,
$\tau_{1}/\tau_{2}$ clearly deviates from 3 and approaches 1. In~\cite{Affouard_prb99}, we have shown that 
this behaviour indicates a crossover to large angle tumbling motions. Similar features have been recently shown
from MD simulations of supercooled liquid $\mathrm{CS_{2}}$~\cite{Kim_jcp04}.
This result is also confirmed from snapshots of single-molecule 
trajectories in Fig.~\ref{figure_snap}. 
Thus, low temperature rotational dynamics can be seen as tumblings between 
prefered orientations along [111] directions of the cubic cell.
This result also proves that simple exponential decay for $C_{l}(t)$ function is not granted 
at low temperature and that 
non-exponential relaxations occur.

\subsection{Collective dynamics}
The coherent intermediate scattering function $S(Q,t)$ has been calculated from MD 
at different temperatures and different $Q$ wave-vectors and it is compared with the NSE data in Fig.~\ref{figure2}
at the arbitrary wave-vector $Q= 1.54 $ \AA$^{-1}$.
A good agreement between 
experimental and numerical $S(Q,t)$ is found. 
However, some 
discrepancies are particularly observed at high temperatures at the longest Fourier times for which 
the NSE $S(Q,t)$ does not decay completely to zero. Since no elastic contribution is 
expected outside the Bragg Peaks, this feature is most likely due to uncertainties 
in the background substraction. At low temperatures, at $T = $ 130 and 140 K, 
difference between experiment and simulation are also seen. Those are also the data which show the largest error bars. 
One explanation could be the use of a different sample (see experimental details) providing a less intense 
scattering signal. 

Upon cooling, $S(Q,t)$ displays a decay classically observed 
in MD simulations~\cite{Chong_pre04} and in NSE experiments of 
glass-forming molecular liquids such as glycerol~\cite{Wutke_jcp96} or CKN~\cite{Mezei_prl87}. 
For DFTCE glassy crystal, the long time relaxation corresponding to the overall rotations
of molecules is separated from short time regime by 
a plateau-like region. This latter can be associated to the 
orientational trapping of molecules, the analogue of the cage effect in liquids~\cite{affouard_jpcm03}, 
as already demonstrated in~\cite{Affouard_prb99}
for others plastic crystals.
In the long time relaxation regime,
we defined the characteristic time $\tau_{1/e}$ of the rotational dynamics 
as the time it takes for $S(Q,t)$ to decay from 1
to $1/e$  at the wave-vector $Q = 1.54$ \AA$^{-1}$. 
This arbitrary definition is chosen in order to fully compare both NSE and MD data. Owing
the limited $Q$ and time domain investigated in the present NSE experiment,
other definitions such as fits with stretched exponential of Kohlrausch or time measured 
to decay from 1 to 0.1 can not be employed.
An alternative definition will be particularly used in the following section concerning the MCT analysis of the MD data.

$\tau_{1/e}$ obtained experimentally and numerically  
are displayed in Fig.~\ref{figure3}. 
We found that both NSE and MD 
characteristic times $\tau_{1/e}$ are consistent with previous 
NMR~\cite{Stokes_jcp79} and dielectric investigations\cite{Kruger_jpcm94} 
and allows us to extend DFTCE data at high temperatures over 3-4 decades. 
It clearly confirms the non-Arrhenius behavior of the rotational dynamics.
For comparison, we also plotted in Fig.~\ref{figure4}, the Vogel-Fulcher-Tammann (VFT) law 
$\tau_{VFT} = \tau_{\infty} \exp [1/K_{VFT}.(T/T_{VFT}-1)]$ where $T_{VFT}$ is the temperature of apparent divergence of $\tau_{VFT}$ and $K_{VFT}$ measures the kinetic fragility. Parameters  $K_{VFT} = 0.13$ and $T_{VFT} = 70$ K were determined
from  dielectric experiments of plastic DFTCE~\cite{Kruger_jpcm94} in the temperature range [95-120] K.
From MD simulation, we obtained $K_{VFT} \simeq $ 0.20  and $T_{VFT} \simeq $ 86.9 K
for $\tau_{1/e}$ in fair agreement 
with the dielectric experiment. 
A perfect agreement can not be expected owing the $Q$-dependence of the neutron scattering technique
and the different temperature domain investigated.

In~\cite{Affouard_prl01,Affouard_prb99}, we suggested that 
some common microscopic mechanisms, relatively well described by MCT, are involved in both 
orientationally disordered crystals and molecular liquids. Therefore, in the following, we
analyse our data in the MCT framework.
In Fig.~\ref{figure5}, $S(Q,t)$ is plotted against the rescaled time $\tau_{1/e}$ previously determined.
We find that one of the  
main predictions of MCT \emph{i.e} the time-temperature superposition principle (TTSP) holds relatively well both for
NSE and MD data. 
We show that the curves overlap for the large time part and collapse gradually onto a master curve upon cooling.
This feature is made particularly clear
since we have developed a realistic MD model of DFTCE which allows us to access data 
in the plateau region of $S(Q,t)$ which was 
not possible in the present NSE experiment. At $T = $130 K, $S(Q,t/\tau_{1/e})$ obtained from MD deviates from the master curve. 
This indicates that MCT ceases to be valid at that temperature. 
This result is confirmed from snapshots and trajectories of target molecules, as already shown for
other plastic crystals in~\cite{Affouard_prb99,Affouard_jncs02}, which reveal
that at the lowest temperatures rotational motions correspond to
activated tumblings of the overall molecules between preferred molecular orientations \emph{i.e} 
the so-called hopping processes not
taken into account in the ideal version
of MCT (see section~\ref{static}). Moreover, this deviation also allows us to roughly estimate the critical temperature $T_{c}$.

Above $T_{c}$, MCT predicts that
the late $\beta$ regime or the early $\alpha$ relaxation
is described by the following power law (going beyond first order):
\begin{equation}
\phi(Q,t) = f^{c}_{Q} - h^{(1)}_{Q}.(t/\tau)^{b}
+ h^{(2)}_{Q}.(t/\tau)^{2b} \label{equation1}
\end{equation}
where $f^{c}_{Q}$ is
the so-called nonergodicity parameter associated to the plateau height and 
$\tau = t_{0}(T-T_{c})^{-\gamma}$ with $\gamma = 1/2a + 1/2b$.
The parameters $t_{0}$,$\gamma$ and $b$ are
temperature and $q$ independent and related via
$\gamma = 1/2a + 1/2b$ and
$\Gamma^{2}(1-a)/\Gamma(1-2a) = \Gamma^{2}(1+b)/\Gamma(1+2b)$
where $\Gamma(x)$ is the gamma function.
The first two terms of Eq.~\ref{equation1} correspond to the classical von Schweidler law and 
the last term is a second order correction.
First,  we fixed the exponent $b$ by performing
individual fits of the different correlators $S(Q,t)$ 
in the late $\beta$ regime using Eq.~\ref{equation1}
at $T=140 $ K expected to be the lowest
temperature where MCT predictions are valid (see Fig.~\ref{figure4}).
We used the following free parameters: $b$, $f^{c}_{Q}$, $\tilde{h}^{(1)} = h^{(1)}(Q).\tau^{-b}$ and
$\tilde{h}^{(2)} = h^{(2)}(Q).\tau^{-2b}$.
Best results are obtained for $b=0.59$ which corresponds to
$a=0.31$ and $\gamma=2.45$.
Then, using this value of $\gamma$, scaling properties in the $\alpha$ relaxation regime have been checked using two functions
$R_{1}(T)$ and $R_{2}(T)$ defined as:
\begin{equation}
R_{1}(T) = \frac{\tau_{Q}^{-1/\gamma}(T) - \tau_{Q}^{-1/\gamma}(T1)}{\tau_{Q}^{-1/\gamma}(T2) - \tau_{Q}^{-1/\gamma}(T1)}(T2-T1) + T1 
\label{R1}
\end{equation}
\begin{equation}
R_{2}(T) = \frac{\tau_{Q}^{-1/\gamma}(T)}{\tau_{Q}^{-1/\gamma}(T2) - \tau_{Q}^{-1/\gamma}(T1)}(T2-T1) 
\label{R2}
\end{equation}
where $T_{1}$ and $T_{2}$ are two arbitrary low temperatures and 
$\tau_{Q}(T)$ is a characteric time of the $\alpha$ relaxation defined as the time  
it takes for $S(Q,t)/S(Q,t=2 \ ps)$ to decay from 1
to $1/e$. This alternative definition eliminates the short-time dynamics and 
ensures that the $\beta$ relaxation does not pertub the $\alpha$-relaxation time~\emph{i.e.}
$\tau_{Q}(T)$ is not influenced by other parameters $f_{Q}$,  $h^{(1)}(Q)$ or $b$. 
Therefore, it 
avoids a tedious fitting procedure with stretched exponential of Kohlrausch.
It is also assumed that any characteristic times
belonging to the  $\alpha$ regime show asymptotically the same
temperature dependence $\tau_{Q} \sim \tau$ defined in Eq.~\ref{equation1}.
For all $Q$ vectors, it can be shown that $R_{1}(T) = T$ and $R_{2}=T-T_{c}$ 
in the temperature range where MCT predictions are
valid. 
In Fig.~\ref{figure5}, $R_{1}$ and $R_{2}$ scaling funtions
are displayed as function of temperature at different $Q$ wave vectors.
They clearly indicate a temperature range [140-180] K where all data obey 
an universal master curve. This result is particularly significative owing 
the wide $Q$-range investigated and confirms the choice for the exponent $b$. 
 Both $R_{1}$ and $R_{2}$ allow us to estimate 
two remarkable temperatures: i) $T_{A} \simeq 180 $ K corresponding to the crossover from 
simple liquids dynamics to  
slow dynamics. 
As observed
in previous MD of molecular glass-forming liquids~\cite{Kaemmerer_pre97},
polymer melts~\cite{Aichele_epjE01} or plastic crystals~\cite{Affouard_prl01}, this temperature
seems very dependent on the chosen correlator $S(Q,t)$~\emph{i.e} on the space range 
probed at the wave-vector $Q$. 
ii) the critical temperature predicted by MCT $T_{c} \simeq 128 $ K. The superposition of the different 
correlators seems better for $R_{1}$ than $R_{2}$. Thus, contrary to MCT predictions, this reveals a weak
$Q$-dependence of $T_{c}$ of about 3 K.
In order to check the limit of this approach,
the same work was also performend on the NSE data and results are displayed as inset
in Fig~\ref{figure5}. It should be mentionned that $\tau_{Q}$ is not accessible in our present experimental data
and we had to use $\tau_{1/e}$ which in intrinsically worse. $\tau_{1/e}$ is influenced by the $\beta$-regime.
 Moreover, $b$ obtained from MD was employed. 
Despite these two drawbacks,  reasonable results are obtained 
for both $R_{1}$ and $R_{2}$ for the narrow $Q$-range investigated in fair agreement with the MD data and clearly mmotivate 
measurement at shorter times using time-of-flight
experiment to describe the plateau regime.

A fundamental property of MCT or MMCT stresses that dynamics are driven by the static density correlator $S(Q)$.
This has been confirmed in several numerical and experimental studies
for simple~\cite{Kob_pre1995} or molecular~\cite{Mossa_pre01,Chong_pre04} glass-forming liquids.
Since a good agreement between experiment and simulation was found for the elastic and the diffuse 
structure factor (see Fig.~\ref{figure1}), we checked the $Q$-dependence
of the $\alpha$-relaxation time $\tau_{Q}$,
the non-ergodicity parameter $f_{Q}$ and the total prefactors 
$\tilde{h}^{(1)}(Q)$ and  
$\tilde{h}^{(2)}(Q)$ at $T = 160 $ K. 
Results are shown in Fig.~\ref{figure6}. 
Clearly, we found that $\tau_{Q}$, $f_{Q}$ oscillate in phase with $S_{d}(Q)$ and $S_{c}(Q)$ while 
the amplitudes 
$\tilde{h}^{(1)}(Q)$ and $\tilde{h}^{(2)}(Q)$ oscillate in phase with $S_{d}(Q)$ and out of phase
with $S_{c}(Q)$. 
It should be noted that the $Q$-modulation
of the non-ergodicity parameter and the total prefactor strongly ressembles to results obtained by Franosch~\emph{et Al.} for a hard-sphere colloidal suspension model~\cite{Franosch_pre97}.

\section{Conclusion}
In this paper, we have investigated static and dynamical properties 
of difluorotetrachloroethane glassy crystal using 
complementary X-ray and NSE experiments
and MD simulations performed for a realistic molecular model.
This study clearly confirms that
supercooled molecular liquids and supercooled orientationally disordered crystal share many properties
in the ps-ns regime where some
precursor features (non-exponentiality or non-Arrhenius behavior) of the glass transition emerge.
We particularly found that 
some 
MCT equations developped for atomic liquids provide a good decription
of 
rotational dynamics:
some MCT predictions such as the superposition
principle or the power law $(T - T_{c})^{-\gamma}$ dependence ($T_{c} \simeq 128 $ K )of the  
$\alpha$-relaxation time caracterizing collective dynamics hold well. 
We also demonstrated that 
the $\alpha$-relaxation time $\tau_{Q}$, 
the non-ergodicity parameter $f_{Q}$ and the amplitudes $h_{1}(Q)$  were 
modulated by the static correlation consistent with the MCT predictions.
In addition, the other remarkable temperature $T_{A}$ was detected at about 
180 K for both self and collective dynamics.
Therefore, in order to fully validate the MD results, it will be of great interest:
i) to complete the present NSE study over a wider $Q$-range and for longer Fourier times 
in order to reach the estimated $T_{c}$ and ii) to make measurement at shorter times using time-of-flight 
experiment to describe the plateau regime.

\section*{Acknowledgments} 

The authors wish to acknowledge the use of the facilities of the IDRIS (Orsay, France)
and the CRI (Villeneuve d'Ascq, France)
where calculations were carried out.
This work was supported by the INTERREG III
program
(Nord Pas de Calais/Kent).


\newpage
\section*{Tables captions}

\begin{table}[htpb]
\caption{Simulation coefficients taken from literature for Buckingham potential
$\phi(r) =  A\exp(-\rho.r) -C/r^{6}$. }
\label{table1}
\end{table}

\begin{table}[htpb]
\caption{Orientational order parameters 
$\langle K_{4} \rangle$, 
$\langle K_{6} \rangle$ and 
$\langle K_{8} \rangle$ at different temperatures.
Values for pure [111], [110] and [100] ordering are also indicated.}
\label{table2}
\end{table}

\newpage
TABLE 1
\bigskip

\begin{tabular}{lccc}\hline
 Site-Site &  $A$ ($\mathrm{kJ mol^{-1}}$)  &  $\rho$ (\AA$^{-1}$)  &
$C$ ($\mathrm{kJ mol^{-1}}$\AA$^{6}$) \\ \hline
C -  C  &   226307   &   0.288  & 2420     \\
C -  F  &   196747   &   0.260  & 1168      \\
C -  Cl  &  390940   &   0.284  & 3864      \\
F -  F  &   171038   &   0.237  &  565     \\
F -  Cl  &   320883  &    0.258  & 1808     \\
Cl -  Cl  &    586389 &     0.284  & 5798   \\ \hline
\end{tabular}

\newpage
TABLE 2
\bigskip

\begin{tabular}{lccc}\hline
Temperature &  $\langle K_{4}\rangle $  &  $\langle K_{6} \rangle$  &  $\langle K_{8} \rangle$ \\ \hline
140 & -0.647  & 0.465 &  0.211 \\
160 &  -0.638 &   0.409 &   0.191 \\
180 &  -0.623 &  0.368 &  0.173 \\
200 & -0.596 &  0.319 &  0.151 \\
220 & -0.577 &  0.291 &  0.134 \\
240 & -0.549 &  0.257 &  0.120 \\
260 & -0.515 &  0.223 &  0.105 \\
$\mathrm{[111]}$ & -1.528 & 2.266 & 0.877 \\
$\mathrm{[110]}$ & -0.573 &-2.071 & 1.665 \\
$\mathrm{[100]}$ &  2.291 & 1.275 & 2.961 \\ \hline
\end{tabular}

\newpage
\section*{Figures captions}
\begin{figure*}[h!]
\caption{\protect\small
Experimental X-ray powder diffraction pattern in the liquid phase at $T=305 $ K(top) and the plastic phase at $T=140$ K (bottom)
as function of the wave-vector $Q$.
The calculated coherent elastic scattering $S_{c}(Q)$ (dotted line) and
the calculated diffuse scattering $S_{d}(Q)$ (dashed line) in the plastic phase are also indicated for comparison.
The DFTCE molecule is displayed in inset.
}
\label{figure1}
\end{figure*}
\begin{figure*}[h!]
\caption{\protect\small
Trajectories (top) of the unit vector $\vec{u}(t) = (x(t),y(t),z(t))$ of
one target molecule over a 1 ns run at 
140 K (left) and 240 K (right). Horizontal lines indicates prefered orientation of $\vec{u}$ along
cubic [111] directions.
Ratio (bottom) of 
reorientational correlation times $\tau_{1}$ 
and $\tau_{2}$ obtained from MD simulation
as function of temperature from 130 to 260 K
The reorientational correlation function $C_{2}(t)$
as  function of time at different temperatures from 130 to 260 K
is given in inset.
}
\label{figure_snap}
\end{figure*}
\begin{figure*}[h!]
\caption{\protect\small
Coherent intermediate scattering function $S(Q,t)$ at $Q = 1.54 \ \mathrm{\AA}^{-1}$
obtained from Neutron Spin Echo experiments (full circle) and MD
computer simulations (solid line) at $T = $ 130, 140, 150, 160, 180 and 200 K.
}
\label{figure2}
\end{figure*}
\begin{figure*}[h!]
\caption{\protect\small
Characteristic time $\tau_{1/e}$ obtained from NSE experiment (full circle) and MD simulation (open circle)
 at $Q = 1.54$ \AA$^{-1}$ as function of the temperature. Data extracted from NMR (square)~\cite{Stokes_jcp79}
and
dielectric experiment using a VFT fit (gray line)~\cite{Kruger_jpcm94} are also indicated.
}
\label{figure3}
\end{figure*}
\begin{figure*}[h!]
\caption{\protect\small
$S(Q,t)$ obtained from NSE experiment at $Q = $ 1.54  \AA$^{-1}$
plotted against the scaled time $t/\tau_{1/e}$
at different temperatures from $T = $ 140 to 190 K.
$S(Q,t)$ calculated from MD simulation at same $Q$ wave-vectors is displayed at different
temperatures from 130 to 220 K.
}
\label{figure4}
\end{figure*}
\begin{figure*}[h!]
\caption{\protect\small
Scaling functions $R_{1}$ and $R_{2}$ at different wave-vectors $Q$
ranging from 0.7 to 3.9 \AA$^{-1}$
as function of the temperature. $R_{1}$ and $R_{2}$ are 
defined in Eq.~\ref{R1} and 
\ref{R2} respectively using exponent $\gamma = 2.45$
extracted from the MCT analysis
in the $\beta$ regime and $T_{1} = 140 $ K and $T_{2} = 170 $ K.
$R_{1}(T) = T$ and  $R_{2}(T) = T - T_{c}$ are expected 
from MCT predictions. $R_{1}$ and $R_{2}$ extracted from NSE experiment
using characteristic times $\tau_{1/e}$, $T_{1} = 150 $ K and $T_{2} = 190 $ K 
are given in inset.
}
\label{figure5}
\end{figure*}
\begin{figure*}[h!]
\caption{\protect\small
$Q$-dependence of different parameters
at the temperature $T = 160 $ K.
a) The $\alpha$-relaxation times $\tau_{Q}$ extracted from the $S_{d}(Q,t)$ correlators. 
b) and c) The non-ergodicity parameter $f_{Q}$ and the total prefator $\tilde{h}^{(1)}(Q)$
obtained from a fitting procedure of $S(Q,t)$
using a von Schweidler law as derived from MCT including second order
corrections (see Eq.~\ref{equation1}).
$\tilde{h}^{(2)}(Q)$ is not shown
but exhibits the same $Q$-dependence as $\tilde{h}^{(1)}(Q)$.
d)
The diffuse scattering $S_{d}(Q)$. First Bragg peaks positions clearly identified from MD
 are also indicated with vertical lines.}
\label{figure6}
\end{figure*}

\newpage
.
\newpage
FIGURE 1

\includegraphics[width=13cm,clip=true]{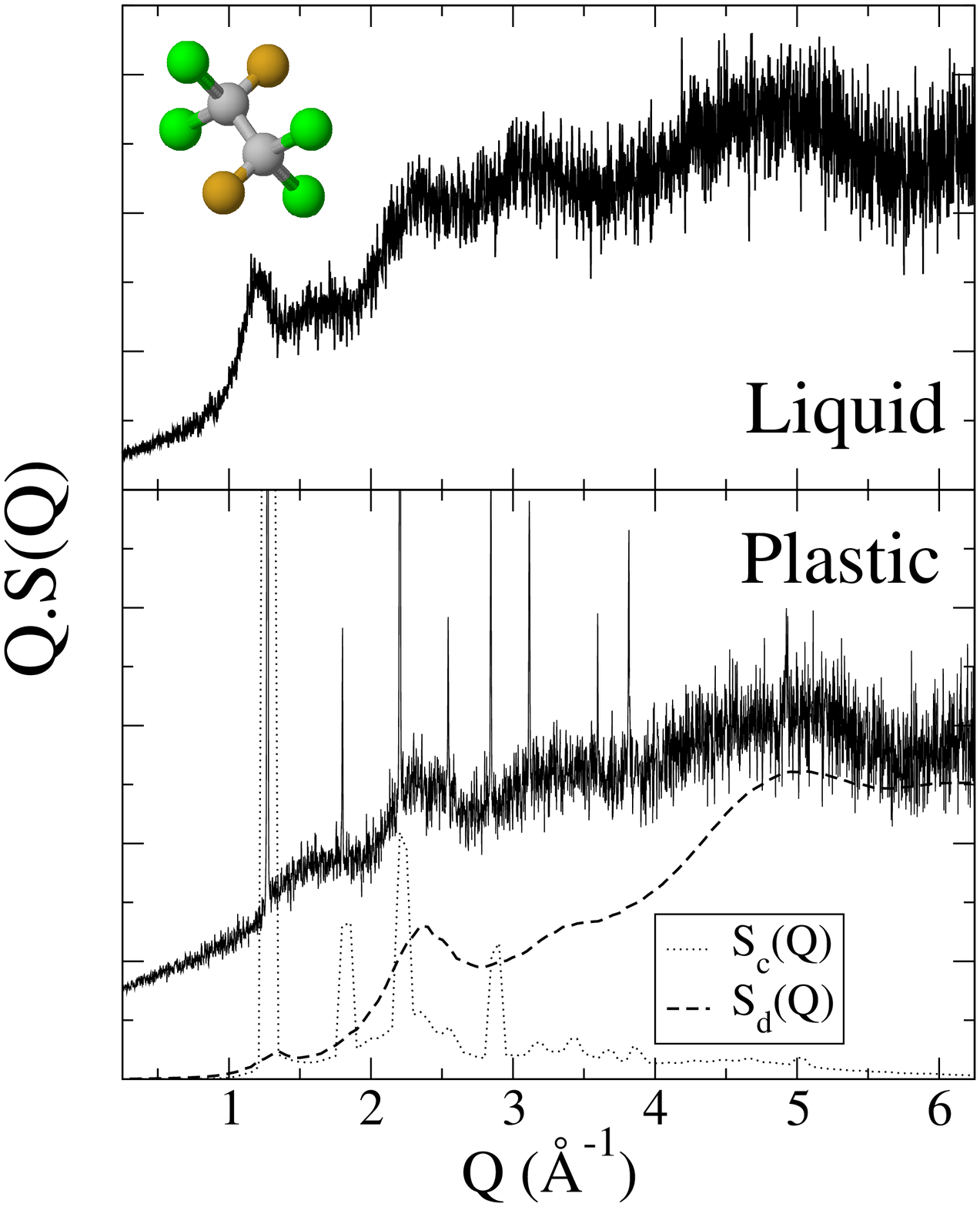}

\newpage
FIGURE 2

\includegraphics[width=12cm,clip=true]{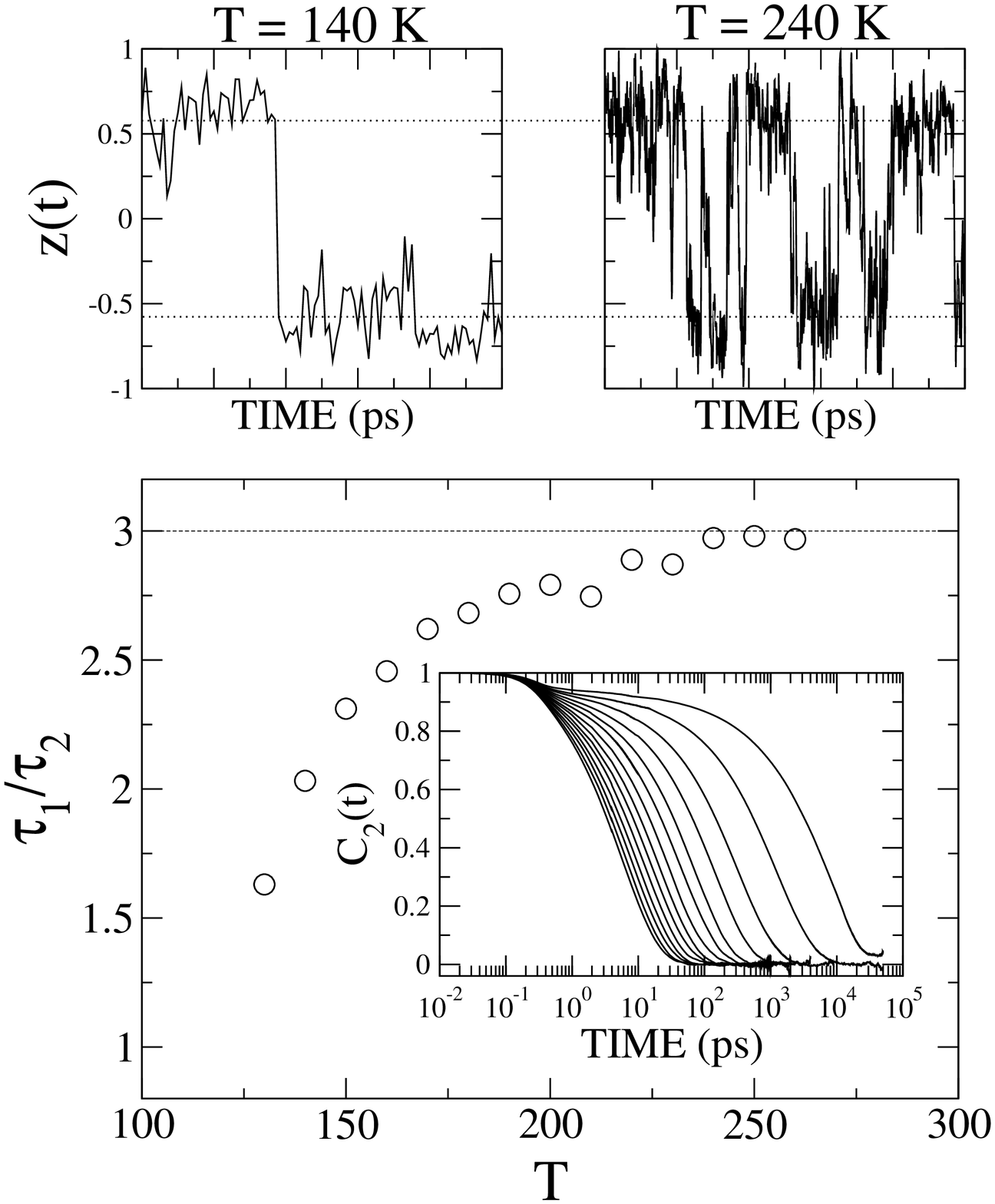}

\newpage
FIGURE 3

\includegraphics[width=8cm,clip=true]{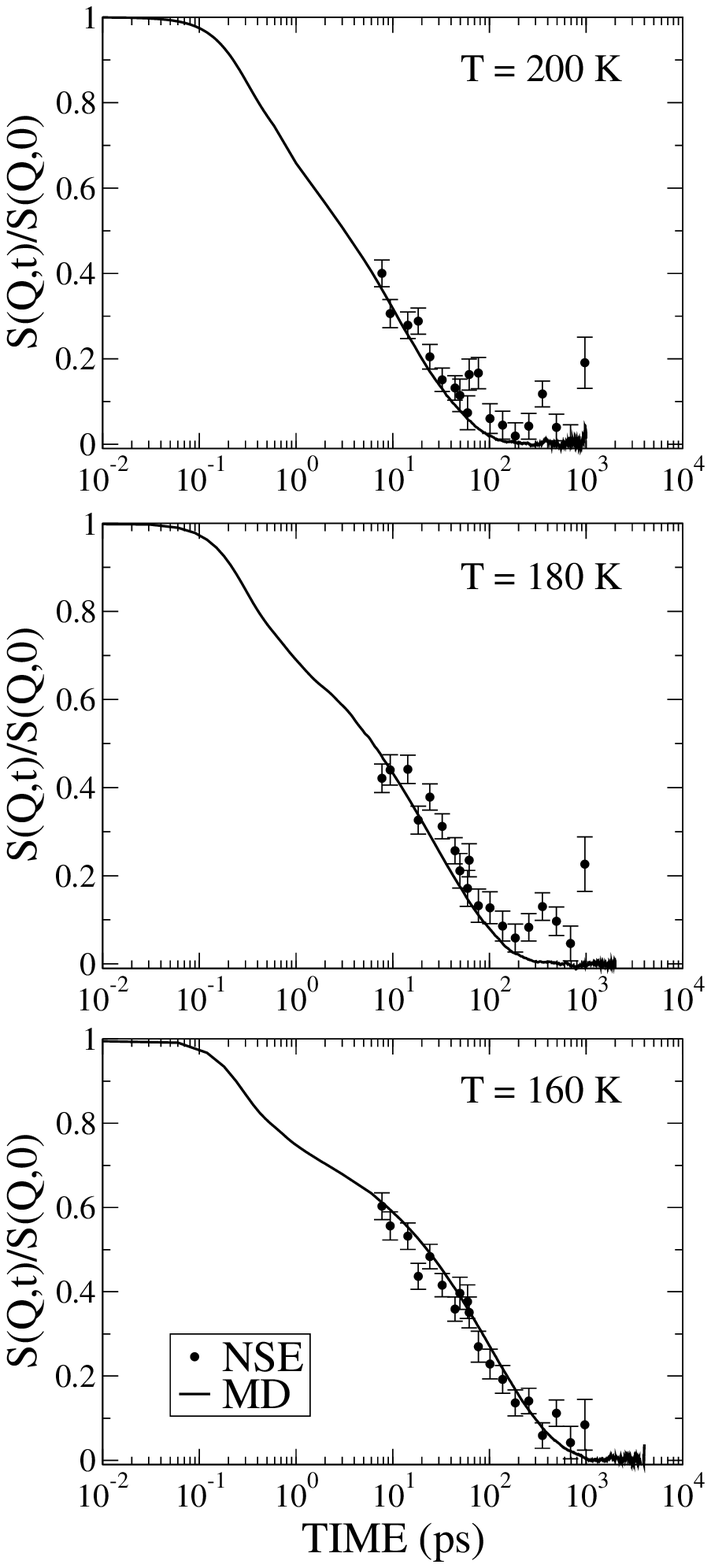}
\includegraphics[width=8cm,clip=true]{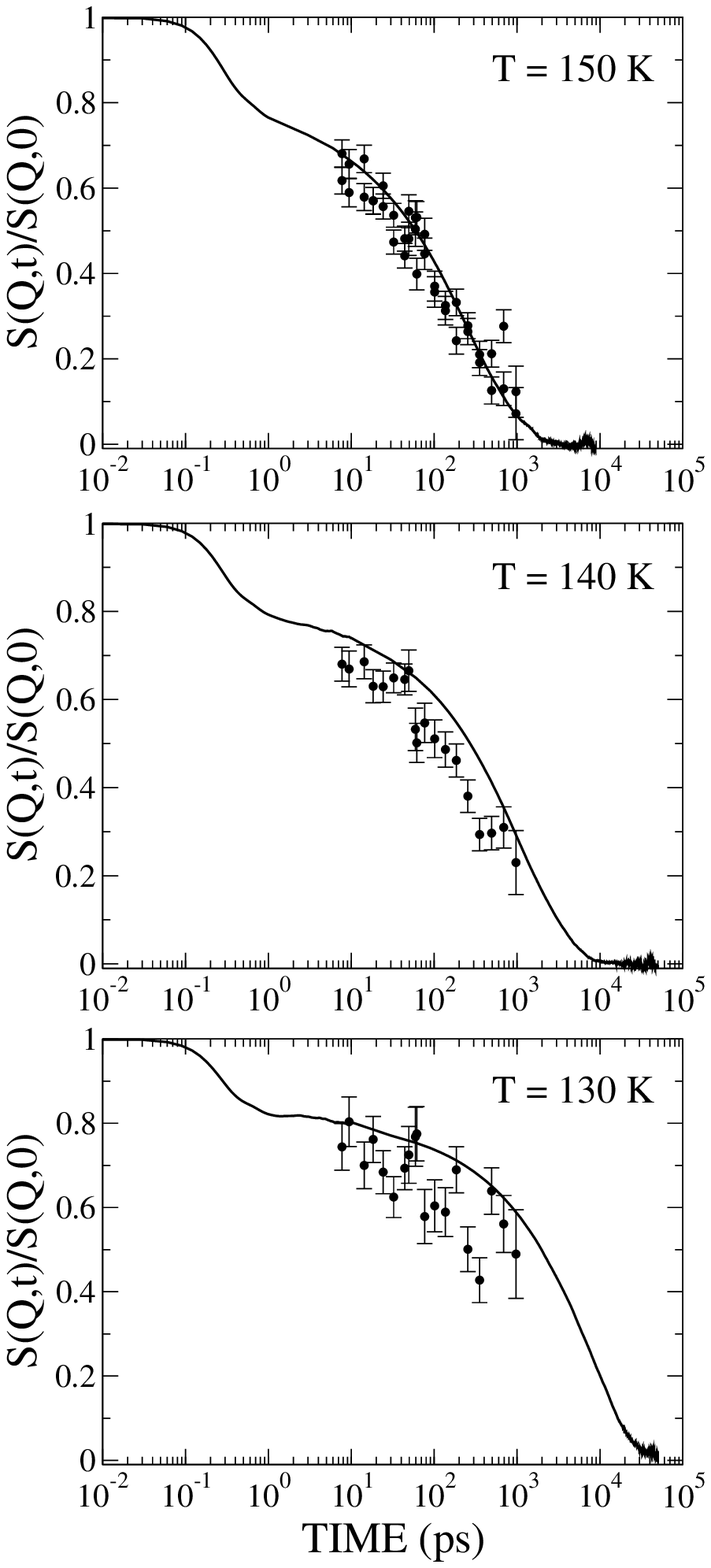}
\newpage
FIGURE 4

\includegraphics[width=12cm,clip=true]{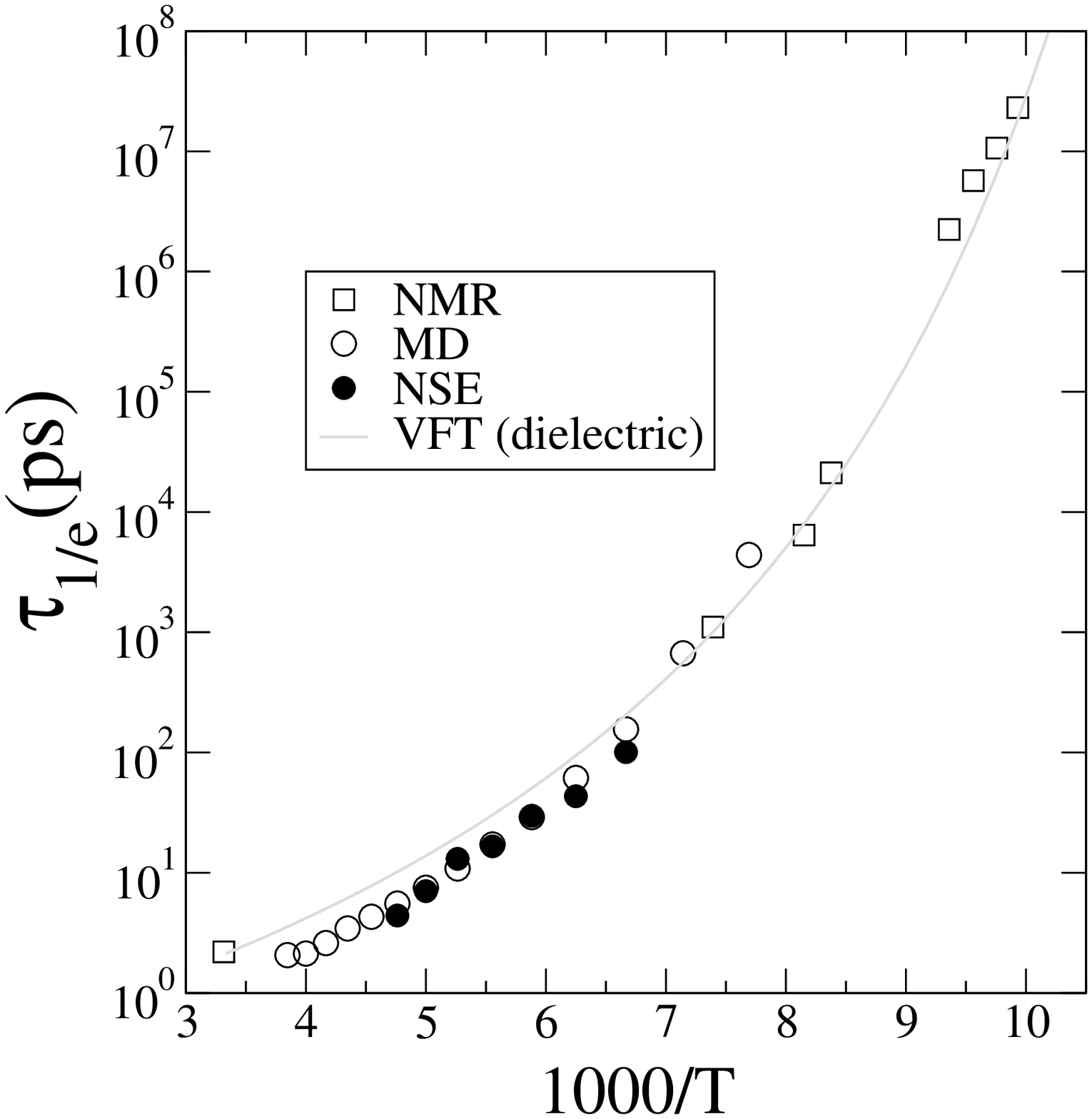}
\newpage
FIGURE 5

\includegraphics[width=12cm,clip=true]{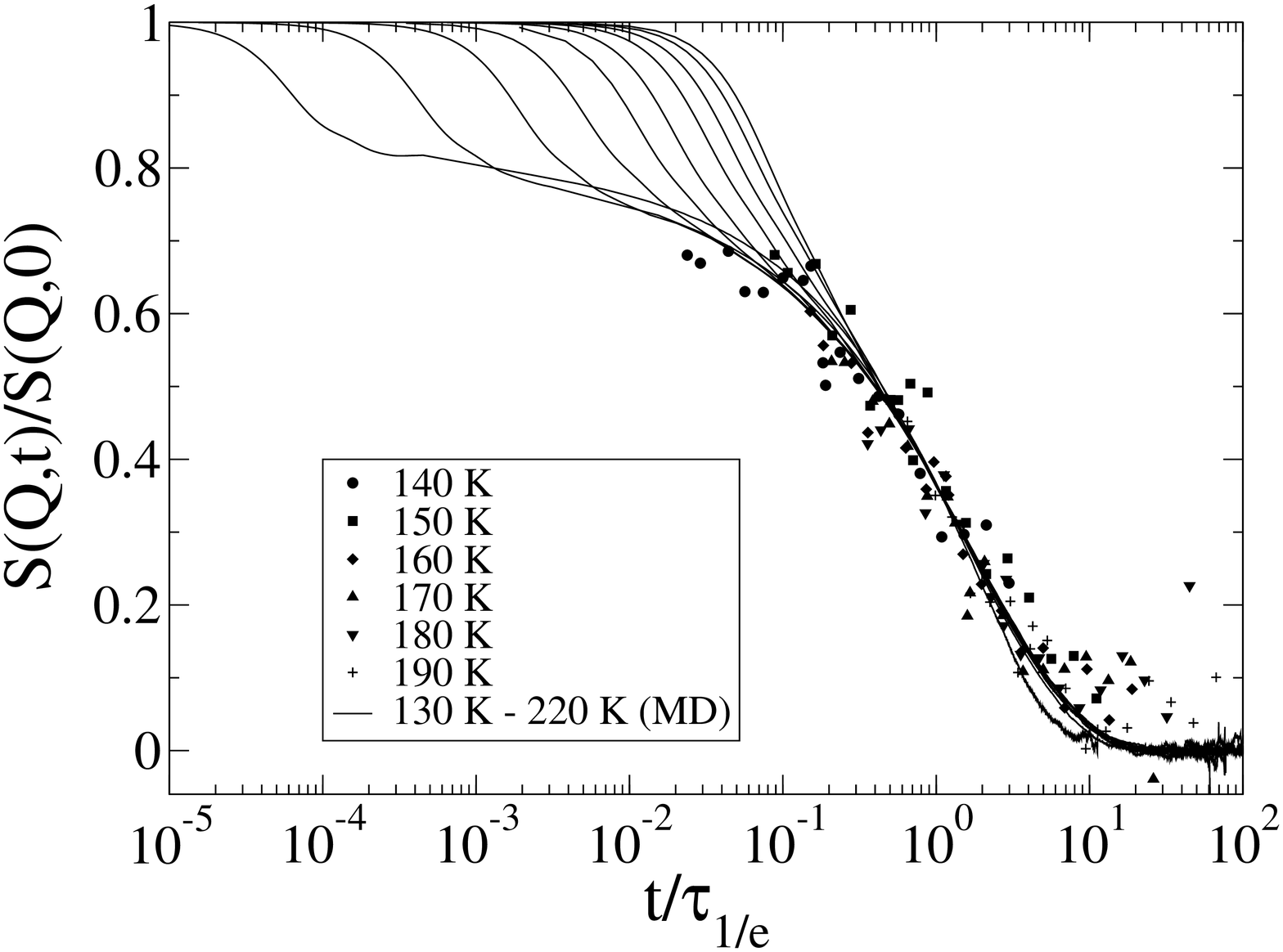}
\newpage
FIGURE 6

\includegraphics[width=11cm,clip=true]{./R1.eps}

\includegraphics[width=12cm,clip=true]{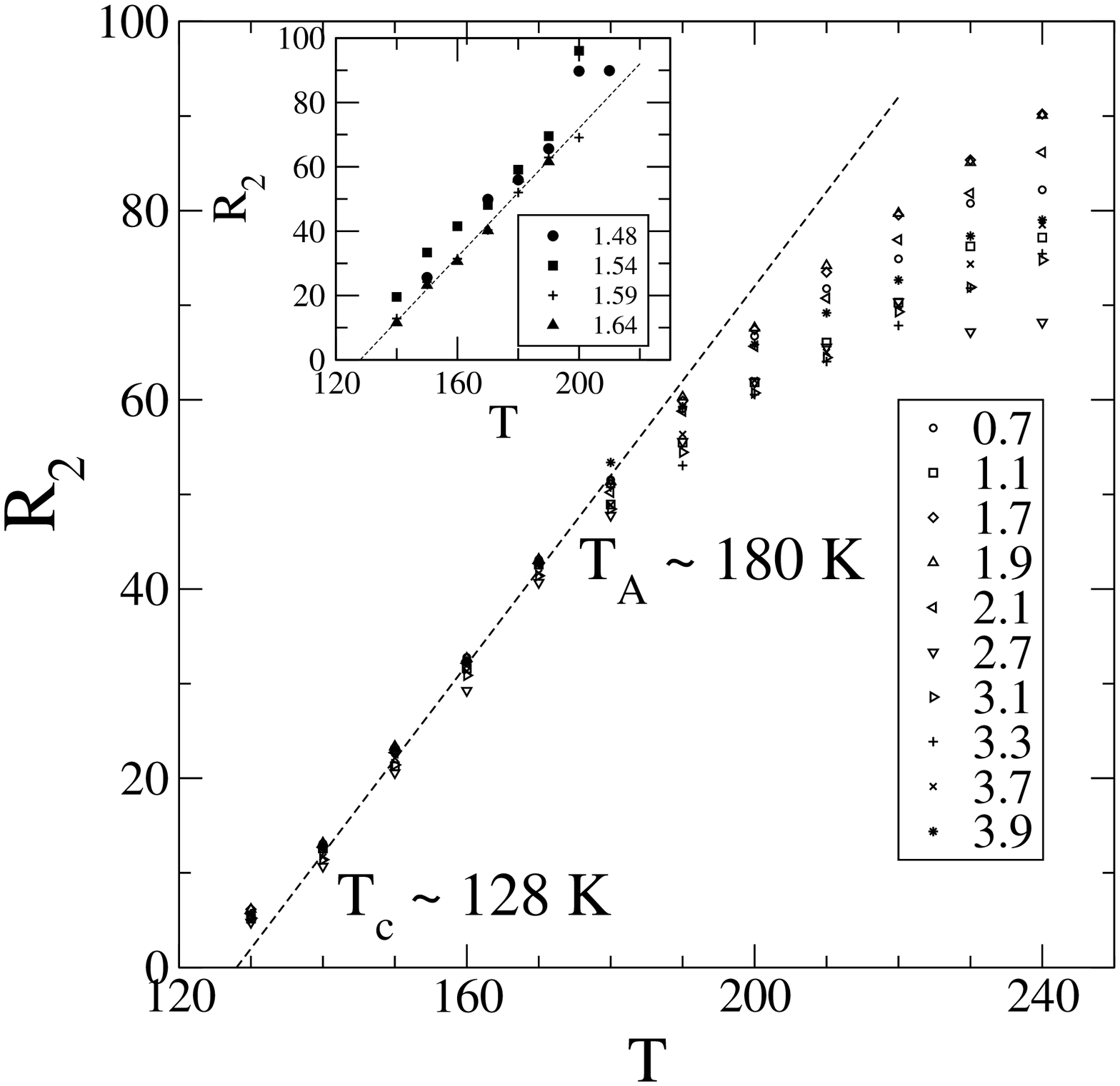}
\newpage
FIGURE 7

\includegraphics[width=12cm,clip=true]{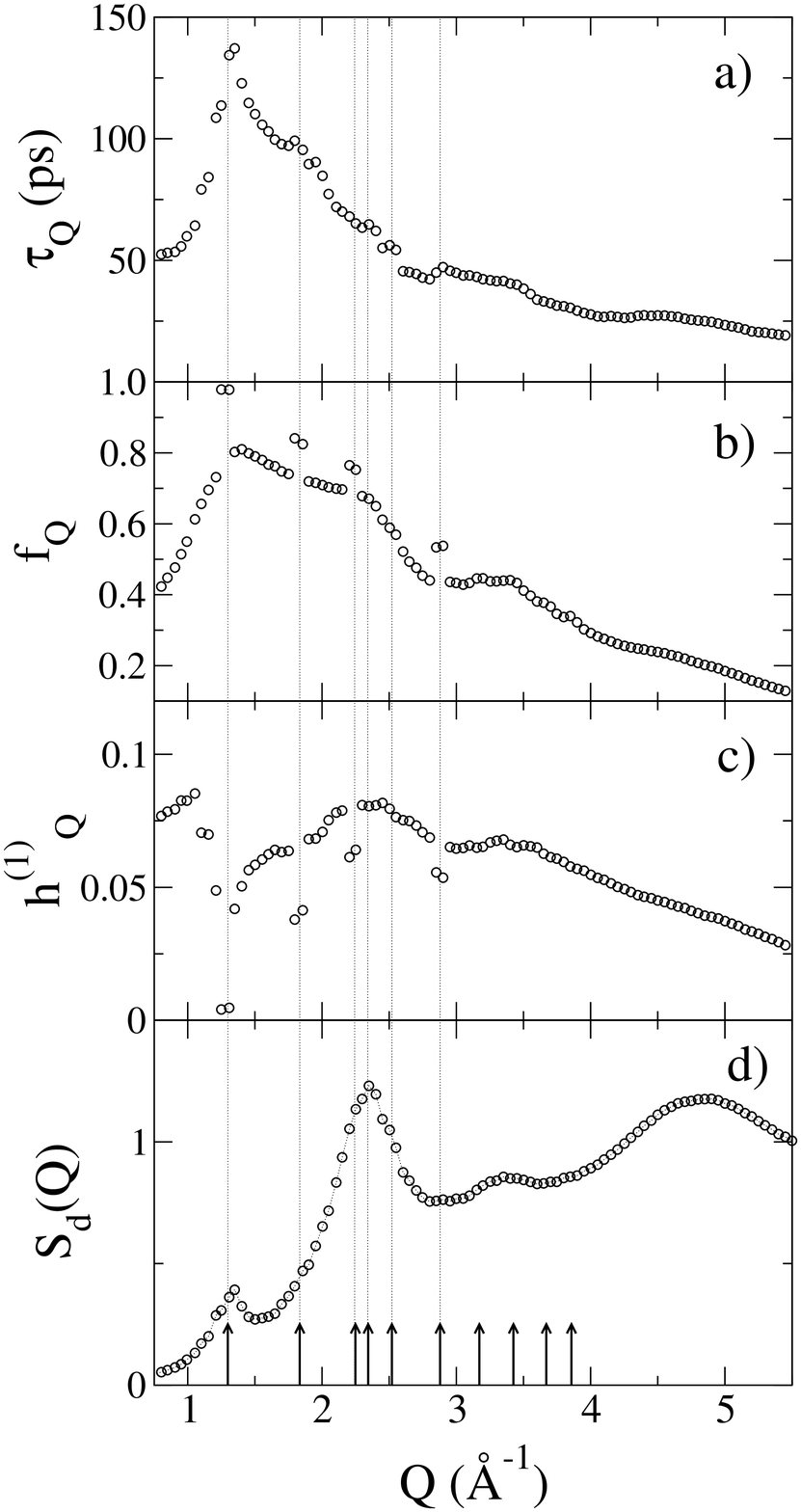}

\end{document}